\documentclass[fleqn,twoside]{article}

\usepackage{graphicx}

\usepackage{amssymb}
\usepackage{amsbsy}

\usepackage{espcrc2}

\newcommand{\be}{\begin{equation}} \newcommand{\ee}{\end{equation}}
\newcommand{\bea}{\begin{eqnarray}} \newcommand{\eea}{\end{eqnarray}}

\newcommand{\re}[1]{(\ref{#1})}

\newcommand{\fig}[1]{Figure \ref{#1}}

\newcommand{\brt}[1]{{\tt #1}}
\newcommand{\rmd}{\textrm{d}}

\newcommand{\Hfour}{H_{\mathrm{4D}}}

\newcommand{\LCDM}{$\Lambda$CDM\ }
\newcommand{\GN}{G_{\mathrm{N}}}

\newcommand{\PRD}[1]{{\it Phys. Rev.} {\bf D#1}}

\newcommand{\PRL}[1]{{\it Phys. Rev. Lett.} {\bf #1}}

\newcommand{\NPB}[1]{{\it Nucl. Phys.} {\bf B#1}}

\newcommand{\PLB}[1]{{\it Phys. Lett.} {\bf B#1}}

\newcommand{\APJ}[1]{{\it Astrophys. J.} {\bf #1}}

\def\lcdm{$\Lambda$CDM\ }
\def\nn{\nonumber}

\title{Cosmological acceleration from a gas of strings}

\author{Francesc Ferrer\address{Physics Department and McDonnell Center for the Space Sciences\\Washington University, St Louis, MO 63130, USA}%
}       
\begin{document}

\begin{abstract}
In string gas cosmology, the extra
dimensions of the underlying theory
are kept at a microscopic scale by a gas of strings.
In the matter-dominated era, however, 
dust pressure can lead to
oscillations of the extra dimensions
and to acceleration in the three visible dimensions, even 
with a vanishing cosmological term. We review the 
resulting oscillating expansion history, that provides an acceptable 
fit to the observed accelerated expansion of the Universe.
\vspace{1pc}
\end{abstract}

\maketitle

\section{Introduction}

The standard $\Lambda$ Cold Dark Matter ($\Lambda$CDM) model, that
reproduces the available experimental observations with remarkable
success, posits that the dynamics of the universe is dominated by dark
matter (DM) and dark energy~\cite{Komatsu:2008hk}. A constant
($\Lambda$) or time-varying (quintessence) vacuum energy could
constitute the dark energy explaining the late-time acceleration in
the expansion rate of the universe. The dominant
component of the matter sector, the DM, does not have appreciable
interactions with radiation, and cannot be in the form of ordinary
baryonic matter as deduced from considerations of cosmological
nucleosynthesis (BBN) together with observations of the anisotropies
in the cosmic microwave background (CMB). In addition, studies of the
dynamics of galaxy clusters show that the DM particles must also be
cold, non-relativistic.

Understanding the
nature of these dark components constitutes one of the 
central challenges in both cosmology and particle physics today.
Indeed, all the particles that have ever been observed or
artificially created in particle colliders so far fall in the 
small 5\% baryonic component (with the exception of neutrinos, that
seem to have too small a mass to play the role of CDM),
although there is no lack of viable candidates for constituting the DM
particles in extensions of the Standard Model (SM) of particle physics.
On the other hand, the observed magnitude of the dark energy (DE)
is orders-of-magnitude smaller than what would be naturally expected.

The particle physics underlying the \lcdm model is partly described at the 
quantum level by the SM, which has been successfully verified by all 
accelerator experiments to date. On the other hand, 
the dynamics of the expansion
is governed in the \lcdm model by Einstein's theory of the 
gravitational interactions. Our present 
knowledge of the gravitational sector at the quantum level, however,
is far from complete, and attempts to further our understanding 
are being actively pursued, particularly in the context of string theory.

Perturbative string theory is most naturally formulated in
9+1 dimensions. The usual way of getting closer to the observed
3+1-dimensional universe is to compactify six spatial dimensions
by hand so as to end up with four-dimensional Minkowski space
times a six-dimensional compact manifold.
The extra dimensions are often taken to be static (indeed,
understanding of string theory in time-dependent backgrounds is
still quite limited), and compactification is
considered not to involve any dynamical evolution.
In the search for a static split into large and small
spatial dimensions, no explanation has emerged for why
there should be three of the former and six of the latter.

From the cosmological point of view, a natural 
possibility is that
the split into three large and six small dimensions
arises due to dynamical evolution.
String gas cosmology (SGC)
(see~\cite{Battefeld:2005,Brandenberger:2008}  
for reviews)
is a cosmological scenario motivated by string theory that, 
unlike in most applications of string theory, treats all
spatial dimensions on an equal footing: they are all
compactified and start out small, and filled with a hot
gas of branes of all allowed dimensionalities.
The branes can wind around the tori. The energy of the winding modes
increases with expansion due to the tension of the branes,
and this resists expansion. As the universe expands and cools
down, winding and anti-winding modes annihilate, allowing further
expansion. A simple counting argument suggests that $p$-branes
and their anti-branes cannot find each other to annihilate in more
than $2p+1$ spatial dimensions, so at most $2p+1$ dimensions
can become large. For $p=1$, corresponding to strings, this is three
spatial dimensions.
Also in contrast to most higher-dimensional proposals,
SGC aims to explain not only why some dimensions are hidden,
but also why the number of visible dimensions is three
(see \cite{Durrer:2005,Karch:2005} for other proposals
along the same lines). 

At late times in the universe, the visible spatial dimensions expand,
while any compact dimensions which exist must be
relatively static. 
Assuming that the dilaton is stabilised by some other mechanism,
the string gas can stabilise the extra dimensions
during the radiation-dominated era. 
However, when the universe becomes matter-dominated,
the matter will push the extra dimensions to open up.
It was shown in \cite{Ferrer:2005} that the gas of strings
can still prevent the extra dimensions from growing too large,
but they cannot be completely stabilised. There is a competition
between the push of matter and the pull of strings.
If the number density of the strings is too small, the extra
dimensions will grow to macroscopic size. If the strings win,
the size of the extra dimensions will undergo damped oscillations
around the self-dual radius.
The oscillations between expansion and contraction of the extra
dimensions induce oscillations in the expansion rate of
the large dimensions, which can involve alternating periods
of acceleration and deceleration \cite{Ferrer:2005}.
(This kind of mechanism has also been studied in
\cite{Perivolaropoulos}.)

Since the oscillations can start only after the universe becomes
matter-dominated, they provide an in-built mechanism for
late-time acceleration in string gas cosmology, one that
alleviates the coincidence problem in a manner similar to
scaling and tracker fields \cite{scaling,tracker}.

However, the oscillating expansion history is quite different
from the \LCDM model which is known to be a good fit to
the observations. A comparison to observations 
of type Ia supernovae (SNe Ia), taking into account the 
BBN constraint on additional radiation degrees of freedom, was 
performed in~\cite{Ferrer:2008}.
In the following, 
we review the string gas model~\cite{Ferrer:2005} and the results, detailed
in~\cite{Ferrer:2008}, showing that 
the oscillating expansion history is not ruled out
by the quality of the fit, although it is disfavoured compared
to the \LCDM model. 


Our scenario is based on ingredients already present in
string gas cosmology and does not require adding new
degrees of freedom or turning on new interactions. The
late-time evolution of the universe is driven 
by (classical) gravitational effects. Also, in contrast
to the \lcdm model, there is a fundamental principle that singles out
the number of observed spatial dimensions.

\section{The string gas model}

In the string gas model discussed in \cite{Ferrer:2005}
the spacetime is ten-dimensional, with the metric
\bea 
  \rmd s^2 =&-& \rmd t^2 + a(t)^2 \sum_{i=1}^3 \rmd x^i\rmd x^i \nn \\
&+& b(t)^2 \sum_{j=1}^6 \rmd x^j\rmd x^j \ ,
\label{metric}
\eea
where $i=1\ldots3$ ($j=1\ldots6$) labels the visible (extra)
dimensions. 
All spatial dimensions are taken to be toroidal,
and $b=1$ corresponds to the self-dual radius given by the string length
$l_s\equiv\sqrt{\alpha'}$. 

We assume that the dilaton has been stabilised 
in a way that leaves the equation of motion of the metric
unconstrained, so that it reduces to the Einstein equation,
$G_{\mu\nu} = \kappa^2 T_{\mu\nu}$. 
$\kappa^2$ is the 10-dimensional gravitational
coupling, and we take the cosmological constant to be zero.

We do not consider additional 
covariantly conserved non-trivial tensors,
besides Einstein's, that can be constructed from the
metric and its first and second derivatives
in more than four dimensions (e.g. the Gauss-Bonnet term). 
These higher order curvature terms can be of
importance in the early universe, and lead to inflation
when there are more than three spatial dimensions.
Inflation terminates if the extra dimensions are stabilised so that
at most three dimensions are free to expand. This relates graceful
exit to the number of large dimensions~\cite{Ferrer:2006}. This
scenario, however, is not realised in the SGC context: 
in an inflating space the string gas will be diluted, and space
isotropizes, with all dimensions growing large.

Given the symmetries of the metric \re{metric}, the energy-momentum
tensor has the form
\bea \label{emt}
  T^{\mu}_{\ \nu} = \textrm{diag}( -\rho(t), 
\overbrace{p(t),\dots}^3,\overbrace{P(t),\dots,P(t)}^6 ) \ ,
\eea
where $p$ and $P$ are the pressure in the visible
and the extra dimensions,  respectively.
\subsection{The matter content.}
In addition to ordinary four-dimensional radiation ($\gamma$)
and pressureless matter ($m$), we have a gas of massless strings ($s$)
with winding and momentum modes in the extra dimensions
and momentum modes in the visible ones.

Assuming that all strings have the same initial momentum in
the visible directions, $M$, the energy-momentum tensor~\cite{Ferrer:2005} 
depends on
four parameters: the scale $M$
and the energy densities $\rho_{\gamma,in}, \rho_{m,in}$ and $\rho_{s,in}$.
The evolution of the system is determined by the
two dimensionless combinations:
\bea \label{rs}
  r \equiv M^{-1} \frac{\rho_{\gamma,in}}{\rho_{m,in}} \quad \quad
  f_s \equiv \frac{\rho_{s,in}}{\rho_{\gamma,in}} \ .
\eea

Rescaling $a\rightarrow M a$, the total energy density reads
\bea \label{rho}
  \rho &=& \rho_{m,in} M^{-3} a^{-3} b^{-6} \left( 1 + r a^{-1} \right.\nn \\
&&\left.+ r f_s \sqrt{ a^{-2} + b^{-2} + b^2 - 2 } \right),
\eea
and the pressures can be written accordingly~\cite{Ferrer:2008}.

\subsection{Oscillations and late-time acceleration.}

The dynamical effects of the gas of strings can be read
from the last term in \re{rho}.
The string gas behaves like a scaling solution \cite{scaling}
in the radiation-dominated era and like a tracker
solution \cite{tracker} in the matter-dominated era
\cite{Ferrer:2005}.
The value $b=1$ is an attractor point: as long as the
initial value of $b$ is not too large ($b<\sqrt{2}$ is a
necessary condition), $b$ will rapidly evolve to unity,
and the extra dimensions are stable.
Then the energy density of the string gas behaves exactly
like radiation.
When the universe becomes matter-dominated, the string gas
starts tracking the matter as the extra dimensions expand.
When the extra dimensions are pulled back and contracted
by the strings, the visible dimensions start oscillating
between deceleration and acceleration.
(If the string gas is too weak to prevent the extra dimensions
from opening up, they will grow without bound,
and there will be no acceleration in the visible dimensions.
We are not interested in this possibility.)

Let us stress that, in our scenario,
the late-time acceleration does not require vacuum energy (indeed,
it would rapidly decompactify the extra dimensions~\cite{Ferrer:2005}).
Since the departure from matter-dominated 4D behaviour is due
to both the extra dimensions and the string gas, the
apparent dark energy does not obey a simple equation of state.
It is noteworthy that the deceleration parameter can dip
below the de Sitter value $-1$. Such rapid acceleration is
usually associated with violation of the null energy condition,
i.e. equations of state more negative than $-1$.
Interestingly, the violation of the null energy
condition by the strings makes it possible
to have acceleration even when the energy density
of the universe is dominated by ordinary matter, with
$\Omega_{tot}\approx\Omega_m<1$.
However, this does not imply spatial curvature, since the
correspondence between spatial flatness and critical density
is broken by the extra-dimensional terms in the Hubble law.

Deep in the radiation-dominated era (in particular, during BBN),
the energy density of the string gas evolves like radiation, and
contributes to the total energy density a fraction
$\Omega_{s,in} = f_s/(1+f_s)$,
given that the contribution of matter is negligible
and $b=1$ in the radiation-dominated era.
The string fraction $f_s$ is related to the effective
number of additional neutrino species $\Delta N_\nu$
by $f_s=7 \Delta N_\nu/43$ \cite{Cyburt:2004}.
Allowing for a large electron neutrino chemical potential, from BBN we have
$\Delta N_\nu\leq4.1$, which translates into
$f_s\leq0.7$, or $\Omega_{s,in}\leq0.4$ \cite{Barger:2003}.
The bound depends on the assumption that the gravitational coupling
during BBN is the same as today, which is not necessarily true
in the string gas model, since $G_N\propto b^{-6}$.
If $b<1$ today, the gravitational coupling at BBN is reduced
relative to the present value, so there is more room for new
degrees of freedom.
However, generally $b$ dips below unity
only very slightly, and typically $b>1$ today, so taking this
into account would make the constraints tighter.
A requirement
for the string gas being able to keep the extra dimensions
small is $r f_s>3/2$~\cite{Ferrer:2005}.
There are no other constraints on $r$, since it depends on $M$,
the initial momentum of the strings in the visible directions,
on which there is no limit.

\section{Constraints from SNe Ia}

The fact that a matter-dominated period followed by
accelerated expansion without decompactification
is possible may be seen as a step towards developing
SGC into a realistic model of the universe at all eras.
However, it is not clear whether the late-time acceleration produced
by this mechanism can be in agreement with observations. 
A detailed study of the parameter space of the model
and a comparison to different cosmological datasets
was undertaken in~\cite{Ferrer:2008}. We
summarise, in the following, the main results of that analysis.

Two important sets of observations which depend only on
the background are luminosity distances of SNe Ia
and the primordial abundance of light elements. 
The Union dataset~\cite{Kowalski:2008} is the newest and most
comprehensive collection of SNe Ia observations, but it has been analysed with
the assumption that the \LCDM model is correct. To check for any
potential bias against models which are
significantly different from $\Lambda$CDM, like the string gas model,
it is convenient to also use the ESSENCE SNIa dataset \cite{Davis:2007}.

In comparing the curve of measured luminosity distances {\it vs.} redshift
of SNe Ia, we should keep in mind that the metric \re{metric} does not
have the FRW form. The usual expression for the luminosity distance does not
hold and the general expression given in~\cite{Syksy:2009} should be used.

\begin{figure}[htb]
\includegraphics[width=0.5\textwidth]{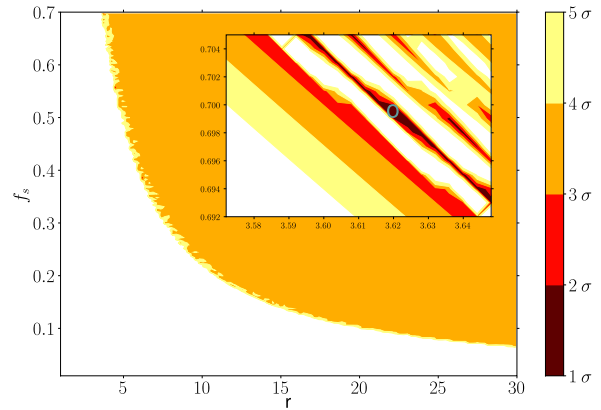}
\caption{Confidence level contours in the $(r,f_s)$-plane for the Union dataset. The best-fit model is marked with a circle.}
\label{fig:CL}
\end{figure}

The results of a scan in the $(r, f_s)$ are shown in \fig{fig:CL}. The
complicated $\chi^2$ contours are not an artifact of the analysis. To
obtain enough acceleration in the visible
dimensions at sufficiently late times, the present day
has to be in a specific location, just after the rise of
one of the first few oscillations. 
Also, in order to have strong acceleration, the extra dimensions
have to expand almost to the point of not turning back,
and then contract rapidly. If the extra dimensions
were to expand slightly more, they would not
turn around, and there would be no acceleration.
Therefore the best fits are obtained on the border
of very poor fits

For the Union dataset, the $\chi^2$ for the best-fit string 
gas model 
without the BBN constraint is 9.3 points worse
than for the \LCDM model, and 21.5 points worse when the
BBN constraint is taken into account~\cite{Ferrer:2008}. 

\begin{figure}[htb]
\includegraphics[width=0.5\textwidth]{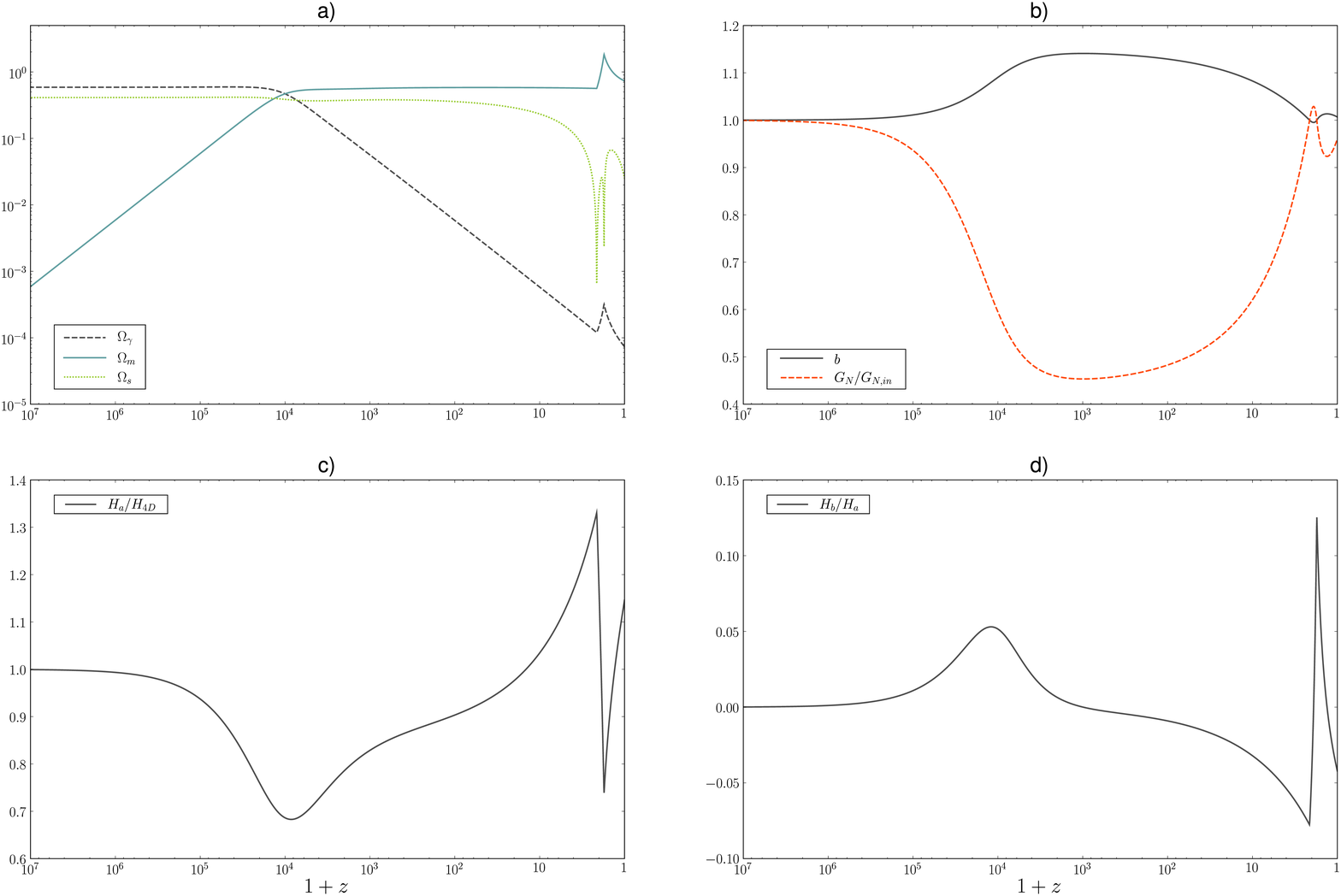}
\caption{a) Density parameters $\Omega_i \equiv \kappa^2\rho_i/(3 H_a^2)$,
b) size of the extra dimensions and Newton's constant,
c) expansion rate of the large dimensions ($H_{4D}$ is the
Hubble parameter in the usual four-dimensional case)
and d) expansion rate of the extra dimensions,
for the best-fit model to the Union data, with the BBN
constraint.}
\label{fig:bestfit}
\end{figure}

In \fig{fig:bestfit}, we plot some quantities for the
best-fit model to the Union dataset (with the BBN constraint
included).
The energy density of the
string gas is completely subdominant at late times, $\Omega_{s0}=0.02$.
However, the string gas can still have a large impact on the dynamics,
because its energy-momentum tensor violates the
null energy condition. When the expansion is faster than in the
Einstein-de Sitter case, the matter density parameter
$\Omega_m\equiv\kappa^2\rho_m/(3 H_a^2)$
is smaller than unity, and in principle it could be in the
observationally allowed range $\Omega_{m0}\approx$ 0.2--0.3 today.
However, for the best-fit model we have $\Omega_{m0}=0.73$,
far too large.

In \fig{fig:bestfit} b) we show the scale factor of the extra
dimensions $b$ and the four-dimensional gravitational coupling
$\GN\propto b^{-6}$. The difference between $b$ at BBN
and today is small, and well within the observational limits
discussed in \cite{Ferrer:2005}. However, $b$ deviates noticeably
from unity at last scattering. This is a generic
feature of the string gas model, because last scattering is soon after
the matter-radiation equality, when the extra dimensions start
opening up. This prediction could provide a stringent constraint.
However, quoted limits on the variation of $\GN$ (or on
new radiation degrees of freedom) from the CMB and other non-BBN
probes are model-dependent, and rely
on perturbation theory. (Note that the
string gas does not behave like radiation at last scattering.)

In \fig{fig:bestfit} c) we show the expansion rate of the
visible dimensions $H_a$ relative to what it would
be without the extra dimensions and the string gas, denoted
by $\Hfour$. 
Comparing to the plot of $H_b/H_a$ in \fig{fig:bestfit} d),
we see how acceleration in the visible dimensions correlates
with contraction of the extra dimensions. 
The Hubble parameter today in the model is somewhat low, which is
related to the large value of $\Omega_{m0}$. 
In order to get enough acceleration in the recent past, it
seems that the extra dimensions must have recently collapsed,
so $b\approx1$ today. The value $\Omega_m=0.3$, for example,
then requires $H_a/\Hfour=1.8$. The maximum value of $H_a/\Hfour$
in the best-fit model is only 1.3, and the value today is 1.1.
Without the BBN constraint, the situation would be better, with
higher values of $H_a/\Hfour$.

The quantity $H_a/\Hfour$ also gives the relation between
the age of the universe and the present value of the Hubble
parameter, since $H_a/\Hfour=3 H_a t/2$ at late times.
A model-independent observational constraint on the
age of the universe is given by the ages of globular
clusters \cite{Krauss:2003}, which lead to the lower limit
$t_0\geq11.2$ Gyr at 95\% C.L. and a best-fit age of $t_0=13.4$ Gyr.
The best model-independent measure of the current
value of the Hubble parameter comes from the Hubble Key
Project \cite{Jackson:2007}. The result is sensitive to the
treatment of Cepheids, and two different analyses yield
$H_{a0}=0.73\pm0.06$ km/s/Mpc and $H_{a0}=0.62\pm0.05$ km/s/Mpc
(1$\sigma$ limits). Taking the best-fit value for $t_0$ and
the mean values for $H_{a0}$ gives $H_a/\Hfour=1.5$
or $H_a/\Hfour=1.27$, respectively. The value in the best-fit
model is too low, but not drastically so, taking into account
the uncertainties in $t_0$ and $H_{a0}$.

\section{Discussion and conclusion.}

While the string gas model does not fit the SNIa data
as well as the \LCDM model, the fit is not decisively worse
when only the SN data are considered, without the BBN constraint.
Nevertheless, even the lower goodness-of-fit corresponds
to a probability of 15\% for the Union data,
which is not enough to rule out the model.
(For comparison, the goodness-of-fit of the \LCDM model
to the first-year WMAP TT data was 3\% \cite{Spergel:2003},
and this was considered strong support for the model.)

As discussed in \cite{Ferrer:2005},
the energy-momentum tensor for the string gas
is expected to be more complex than \re{rho},
which assumes that all strings have the same momentum
$M l_s/a$ in the visible dimensions. Since 
the evolution is very sensitive to the parameters of the string gas,
a more realistic distribution of strings
with different momenta, 
will lead to quantitatively slightly different oscillations.
To explore this possibility, we would have to know
the distribution of string  momenta, which depends on how the
string gas was created in the early universe and whether it has
thermalised.

The string gas cosmology context aside, this provides
an interesting demonstration of how a model with an expansion
history radically different from \LCDM, but which still
provides a good fit to the supernova data.
In this context, it may be interesting that the Hubble parameter
inferred from observations of the ages of passively evolving
galaxies shows oscillations \cite{ages}, though
it is premature to draw strong conclusions from the data.

\section*{Acknowledgments}

FF is supported by grants from the DOE and NSF.

\end{document}